\documentclass[aps,twocolumn]{revtex4}
\usepackage{graphicx}
\usepackage{dcolumn}
\usepackage{amsmath}
\def\Journal#1#2#3#4{{#1} {\bf#2}, #3 (#4)}

\def\NPB{{\rm Nucl. Phys.} B}
\def\PLB{{\rm Phys. Lett.}  B}
\def\PRL{\rm Phys. Rev. Lett.}
\def\PRD{{\rm Phys. Rev.} D}

\def\ZPC{{\rm Z. Phys.} C}
\def\JPG{{\rm J. Phys.} G}
\def\EPJC{{\rm Eur.Phys.J.}C}
\def\IJMPA{{\rm Int. J. Mod. Phys.} A}

\def\ep{\epsilon}
\def\lam{\lambda}

\def\la{\langle}
\def\ra{\rangle}

\def\al{\alpha}

\def\be{\begin{equation}}
\def\ee{\end{equation}}
\def\bea{\begin{eqnarray}}
\def\eea{\end{eqnarray}}
\begin{document}
\title{Non-leptonic two-body decays of the
$B_c$ meson in light-front quark model and QCD factorization
approach}
\author{ Ho-Meoyng Choi$^{a}$ and Chueng-Ryong Ji$^{b}$\\
$^a$ Department of Physics, Teachers College, Kyungpook National
University,
     Daegu, Korea 702-701\\
$^b$ Department of Physics, North Carolina State University,
Raleigh, NC 27695-8202}
\begin{abstract}
 We study exclusive nonleptonic two-body
$B_c\to(D_{(s)},\eta_c,B_{(s)})+F$ decays with $F$(pseudoscalar or
vector meson) factored out in the QCD factorization approach.
The nonleptonic decay amplitudes are related to the product of
meson decay constants and the form factors for semileptonic $B_c$
decays. As inputs in obtaining the branching ratios for a large
set of nonleptonic $B_c$ decays, we use the weak form factors
for the semileptonic $B_c\to(D_{(s)},\eta_c,B_{(s)})$ decays
in the whole kinematical region and the unmeasured meson decay
constants obtained from our previous light-front quark model.
We compare our results for the branching ratios with those of
other theoretical studies.
\end{abstract}


\maketitle
\section{Introduction}
The discovery of the $B_c$ meson by the Collider Detector at
Fermilab(CDF) Collaboration~\cite{CDF98} in $p\bar{p}$ collisions
at $\sqrt{s}=1.8$ TeV and the subsequent measurement of its
lifetime have provided a new window for the analysis of the heavy-quark
dynamics and thus for an important test of quantum chromodynamics.
Recently the CDF and D0 Collaborations announced some new measurements of the
$B_c$ meson lifetime and mass~\cite{CDF06,D008},
$\tau_{B_c}=0.463^{+0.073}_{-0.065}(\rm stat)\pm 0.036(\rm syst)$
ps~~\cite{CDF06}, $M_{B_c}=6275.6\pm 2.9(\rm stat)\pm 2.5(\rm
syst)$ MeV~\cite{CDF06}, and $6300\pm 14(\rm stat)\pm 5(\rm syst)$
MeV~\cite{D008}. The LHC is expected to produce around $\sim
5\times 10^{10}$ $B_c$ events per year~\cite{Bc10,Gouz}. This will
provide more detailed information on the decay properties of the
$B_c$ meson.
Since the $B_c$ mesons carry flavor explicitly($b$ and $c$) and
cannot annihilate into gluons, they are stable against strong and
electromagnetic annihilation processes. The decays of the $B_c$ meson are
therefore only via weak interactions, which can be divided into
three classes at the
quark level: (1) the $b\to q$ ($q=c,u$) transition with the $c$ quark
being a spectator, (2) the $c\to q$ ($q=s,d$) transition with the $b$
quark being a spectator, and (3) the weak annihilation channels.
Although the phase space of the $c\to s,d$ transitions is much smaller
than the phase space of the $b\to c,u$ transitions, the
Cabibbo-Kobayashi-Maskawa(CKM) matrix elements are greatly in favor of the
$c$ quark decay, i.e.$|V_{cb}|<<|V_{cs}|$. In fact, the
$c$-quark decays provide about $\sim 70\%$ of the $B_c$ decay width
while the $b$-quark decays and the weak annihilation yield about
20$\%$ and 10$\%$, respectively~\cite{Gouz}. This indicates that both
$b$-and $c$-quark decay processes contribute to the $B_c$ decay width
on a comparable footing.

Because the $b$ and $c$ quarks can decay individually and the $B_c$ meson has
a sufficiently large mass, one can study a great variety of decay channels.
There have been many theoretical efforts to calculate the
semileptonic~\cite{Gouz,
CNP,KKL,HZ,IKS01,IKS05,IKS06,EFG67,EFG03D,EFG03E,CC94,LC97,AMV,CD00,NW,HNV,
WSL,LM,DW,God,DSV,AKN} and
non-leptonic~\cite{Bc10,Gouz,KKL,IKS06,EFG03D,EFG03E,CC94,LC97,AMV,CD00,HNV,DW,
DSV,DSV2,Giri,KisJPG,LL08,CDL,FKS04,FW00,KPS,CMM,VAS,DD99,Mas92,WK92,SYDM,IKP,HS84}
decays of the $B_c$ meson. The semileptonic $B_c$ decays provide a
good opportunity to measure not only the CKM elements such as $|V_{cb}|$,
$|V_{ub}|$,$|V_{cs}|$ and $|V_{cd}|$ but also the weak form factors for the
transitions of $B_c$ to bottom and charmed mesons. The
nonleptonic $B_c$ decays, in which only hadrons appear in the
final state, are strongly influenced by the confining color forces
among the quarks. While in the semileptonic transitions the
long-distance QCD effects are described by a few hadronic form
factors parametrizing the hadronic matrix elements of quark
currents, the nonleptonic processes are complicated by the phenomenon
of the quark rearrangement due to the exchange of soft and hard
gluons. The theoretical description of the nonleptonic decays
involves the matrix elements of the local four-quark operators.
Although the four-quark operators are more complicate than
the current operators involved in the semileptonic decays,
the nonleptonic decays of the heavy mesons are useful for
exploring the most interesting aspect of the QCD, i.e. its
nonperturbative long-range character.

In our recent paper~\cite{CJBc}, we analyzed the semileptonic
$B_c$ decays such as $B_c\to (D,\eta_c,B,B_s)\ell\nu_\ell$ and
$\eta_b\to B_c\ell\nu_\ell(\ell=e,\mu,\tau)$ using our light-front
quark model(LFQM) based on the QCD-motivated effective LF
Hamiltonian~\cite{CJ1,CJ2,JC,CJK02,Choi07,Choi08}.
The weak form factors $f_{\pm}(q^2)$ for the semileptonic decays
between two pseudoscalar mesons are obtained in the $q^+=0$ frame
($q^2=-{\bf q}^2_\perp<0$) and then analytically continued to the timelike
region by changing ${\bf q}^2_\perp$ to $-q^2$ in the form factor.
The covariance (i.e., frame independence) of our model has been checked by performing the
LF calculation in the $q^+=0$ frame in parallel with the manifestly
covariant calculation using the exactly solvable covariant fermion
field theory model in $(3+1)$ dimensions. We also found the zero-mode contribution
to the form factor $f_-(q^2)$ and identified the zero-mode operator that
is convoluted with the initial and final state LF wave functions.

In this paper, we extend our previous LFQM analysis of the
semileptonic $B_c$ decays~\cite{CJBc} to the nonleptonic two-body
decays of $B_c$ mesons such as $B_c\to (D_{(s)},\eta_c,B_{(s)})P$
and $B_c\to (D,\eta_c,B_{(s)})V$(here $P$ and $V$ denote
pseudoscalar and vector mesons, respectively).
The QCD factorization approach is widely used since it works
reasonably well in heavy-quark
physics~\cite{BSW,NB97,Cheng,SHB,DGS,Kamal86}. The factorization
approximates the complicated non-leptonic decay
amplitude into the product of the meson decay constant and the form factor.
A justification of this assumption is usually based on the idea of color
transparency~\cite{Bj}. We shall use the form factors for semileptonic
$B_c\to (D_{(s)},\eta_c,B_{(s)})$ decays as well as the meson decay
constants obtained in our LFQM~\cite{CJBc,Choi07} as input
parameters for the nonleptonic $B_c$ decays. As done by many
others~\cite{Gouz,KKL,IKS06,EFG03D,EFG03E,CC94,LC97,AMV,CD00,HNV},
we consider only the contribution of current-current operators
at the tree level and calculate the decay widths for various
nonleptonic $B_c$ decays. As far as the decay width is concerned, the
contribution from the tree diagram is much larger than that from the
penguin diagram. The penguin contribution may be
important in evaluating the $CP$ violation and looking for new physics beyond
the standard model, which we do not consider in this work.

The paper is organized as follows. In Sec. II, we
discuss the weak Hamiltonian responsible for the nonleptonic
two-body decays of the $B_c$ meson. In Sec. III, we present the input
parameters such as the weak decay constants and the form factors
obtained in our LFQM~\cite{CJBc,Choi07} based on the QCD-motivated
effective Hamiltonian~\cite{CJ1,CJ2}. The mixing angles
between $\eta$ and $\eta'$ mesons are also analyzed, both in
octet-singlet and quark-flavor bases, to extract the decay
constants relevant to $\eta$ and $\eta'$ mesons. Section IV is devoted to the
numerical results. A summary and conclusions follow in Sec.V.

\section{Nonleptonic two-body decays of the $B_c$ meson}
The nonleptonic weak decays are described in the
standard model by a single $W$ boson exchange diagram at tree level.
In the standard model, the nonleptonic $B_c$ decays are described by the
effective Hamiltonian, which was obtained by integrating out the heavy
$W$ boson and top quark. For the case of $b\to c,u$ and $c\to s,d$
transitions at the quark level, neglecting QCD penguin operators,
one gets the following effective weak Hamiltonian:
 \bea\label{Hbcu}
 {\cal H}^{b\to c(u)}_{\rm
eff}&=&\frac{G_F}{\sqrt{2}}\biggl\{ V_{cb}[c_1(\mu){\cal O}^{cb}_1
+
c_2(\mu){\cal O}^{cb}_2]\nonumber\\
&&+ V_{ub}[c_1(\mu){\cal O}^{ub}_1 + c_2(\mu){\cal O}^{ub}_2] +
{\rm h.c.} \biggr\},
 \eea
and
 \bea\label{Hcsd}
 {\cal H}^{c\to s(d)}_{\rm
eff}&=&\frac{G_F}{\sqrt{2}}\biggl\{ V_{cd}[c_1(\mu){\cal O}^{cd}_1
+
c_2(\mu){\cal O}^{cd}_2]\nonumber\\
&&+ V_{cs}[c_1(\mu){\cal O}^{cs}_1 + c_2(\mu){\cal O}^{cs}_2] +
{\rm h.c.} \biggr\},
 \eea
where $G_F$ is the Fermi coupling constant and $V_{q_1q_2}$ are the
corresponding CKM matrix elements. We use the central values of
the CKM matrix elements quoted by the Particle Data
Group(PDG)~\cite{Data08} that we summarize in Table~\ref{t1n}.
\begin{table}
\caption{Values for CKM matrix elements used in this work.
}\label{t1n}
\begin{tabular}{cccccc} \hline\hline
$V_{ud}$\;\;\;\;\;\;& $V_{us}$ \;\;\;\;\;\;& $V_{cd}$
\;\;\;\;\;\;&
$V_{cs}$ \;\;\;\;\;\;& $V_{cb}$\;\;\;\; \;\;& $V_{ub}$\\
\hline 0.974\;\;\;\;\;\; & 0.2255\;\;\;\;\;\; & -0.230\;\;\;\;\;\;
& 1.04\;\;\;\;\;\;
& 0.0412\;\;\;\;\;\; & 0.00393 \\
 \hline\hline
\end{tabular}
\end{table}
The effective weak Hamiltonian consists of products of local
four-quark operators ${\cal O}_{1,2}$ renormalized at the scale
$\mu$, and scale-dependent Wilson coefficients $c_{1,2}(\mu)$,
which incorporate the short-distance effects arising from the
renormalization of ${\cal H}_{\rm eff}$ from $\mu=m_W$ to
$\mu=O(m_b)$.
   The
local four-quark operators ${\cal O}_1$ and ${\cal O}_2$ due to
$b$ and $c$ decays are given by
 \bea\label{OI}
{\cal O}^{qb}_1&=& (\bar{q}b)_{V-A}[(\bar{d'}u)_{V-A}+(\bar{s'}c)_{V-A}](q=c,u), \nonumber\\
{\cal O}^{qb}_2&=& (\bar{q}u)_{V-A}(\bar{d'}b)_{V-A} +
(\bar{q}c)_{V-A}(\bar{s'}b)_{V-A}(q=c,u),\nonumber\\
{\cal O}^{cq}_1&=& (\bar{c}q)_{V-A}(\bar{d'}u)_{V-A}(q=s,d), \nonumber\\
{\cal O}^{cq}_2&=& (\bar{c}u)_{V-A}(\bar{d'}q)_{V-A}(q=s,d),
 \eea
 where $(\bar{q}q)_{V-A}=\bar{q}\gamma_\mu(1-\gamma_5)q$ and
 the rotated antiquark fields are given by
 \bea\label{aq}
 \bar{d'}&=&V_{ud}\bar{d} + V_{us}\bar{s},
 \nonumber\\
 \bar{s'}&=&V_{cd}\bar{d} + V_{cs}\bar{s}.
  \eea
Without strong-interaction effects, one would have $c_1=1$
and $c_2=0$. However, this simple result is modified by gluon
exchange: i.e., the original weak vertices get renormalized and
the new types of interactions(such as the operators ${\cal O}_2$)
are induced~\cite{NB97}.
In these decays, the final hadrons are
produced in the form of pointlike color-singlet objects with a
large relative momentum.
Thus, the hadronization of the decay products occurs after
they separate far away from each other.
This provides the possibility to avoid the final
state interaction. A more general treatment of factorization
was presented in~\cite{BBNS,BS}.

For the operators ${\cal O}_1=(\bar{q}_1 q_2)_{V-A}(\bar{q'}_1
q'_2)_{V-A}$ and ${\cal O}_2=(\bar{q}_1 q'_2)_{V-A}(\bar{q'}_1
q_2)_{V-A}$, using the Fierz transformation under which $V-A$
currents remain $V-A$ currents, one gets the following equivalent
forms
 \be\label{Fierz}
c_1{\cal O}_1 + c_2{\cal O}_2= a_{1}{\cal O}_{1} +
c_{2}\tilde{\cal O}_{2}= a_{2}{\cal O}_{2} + c_{1}\tilde{\cal
O}_{1},
 \ee
where
 \be\label{a12}
 a_1(\mu) = c_1(\mu) +\frac{1}{N_c}c_2(\mu),\;\;
 a_2(\mu) = c_2(\mu) + \frac{1}{N_c}c_1(\mu),
 \ee
 and $N_c$ is the number of colors. The terms
 $\tilde{\cal O}_1=(\bar{q}_1 T^aq'_2)_{V-A}(\bar{q'}_1 T^a
q_2)_{V-A}$ and
 $\tilde{\cal O}_2=(\bar{q}_1 T^a q_2)_{V-A}(\bar{q'}_1 T^a
q'_2)_{V-A}$ with SU(3) color generators $T^a$ are the
nonfactorizable color-octet current operators, which are neglected
in the factorization assumption. A detailed analysis of $1/N_c$
corrections to the coefficients $a_1,a_2$ as well as the role of
color-octet current operators in $B$ decays can be found
in~\cite{NB97}.

 In the factorization approach to nonleptonic meson decays, one can
 distinguish three classes of decays for which the amplitudes have the following
 general structure~\cite{BSW}:
 \bea\label{123}
 ({\rm class\; I})&:&\frac{G_F}{\sqrt{2}}V_{CKM} a_1(\mu)\la{\cal
 O}_1\ra_F,\\
 ({\rm class\; II})&:&\frac{G_F}{\sqrt{2}}V_{CKM} a_2(\mu)\la{\cal
 O}_2\ra_F,\\
 ({\rm class\; III})&:&\frac{G_F}{\sqrt{2}}V_{CKM} [a_1(\mu) + x a_2(\mu)]\la{\cal
 O}_1\ra_F,
 \eea
where $\la{\cal O}_i\ra_F$ represents the hadronic matrix element
given as the products of matrix elements of quark currents and $x$ is
a nonperturbative factor equal to unity in the flavor symmetry
limit~\cite{NB97}. The first (second) class is caused by a
color-favored (color-suppressed) tree diagram and contains those
decays in which only a charged (neutral) meson can be generated
directly from a color-singlet current.
The first and second class decay amplitudes are proportional to $a_1$ and
$a_2$, respectively. The third class of
transitions consists of those decays in which both $a_1$ and $a_2$ amplitudes
interfere.

In this paper, we consider the following type of nonleptonic
$B_c\to F_1 + F_2$ where $F_1$ is the pseudoscalar meson($\eta_c,
D_{(s)}, B_{(s)}$) and $F_2$ the meson(vector or pseudoscalar)
being factored out. For instance, the factorized matrix element of
$B^+_c\to F_1F^+_2$ with $F_1=\eta_c$ and $F_2=\pi$  is defined
as
 \be
 X^{(B^+_cF_1,F^+_2)}\equiv
 \la F_1|(\bar{c}b)_{V-A}|B_c\ra
   \la F_2|(\bar{d}u)_{V-A}|0\ra.
   \ee
The matrix elements of the semileptonic $B_c\to F_1$ decays
can be parametrized by two Lorentz-invariant form
factors:
 \be\label{PPF1}
\la F_1(P_2)|(\bar{q}q')_{V-A}|B_c(P_1)\ra
= f_{+}(q^2)P^{\mu} + f_-(q^2)q^\mu,
 \ee
  where
$P=P_1 + P_2$ and $q=P_1-P_2$. The two form factors also satisfy
the following relation:
 \bea\label{FP0M1}
f_0(q^2) = f_+(q^2) + \frac{q^2}{M^2_1-M^2_2}f_-(q^2).
 \eea
The decay constants $f_P$ and $f_V$ of pseudoscalar($P$) and
vector($V$) mesons are defined by
 \bea\label{fpv1}
 \la P(p)|(\bar{q}q')_{V-A}|0\ra&=&-if_P p^\mu,
\nonumber\\
\la V(p,h)|(\bar{q}q')_{V-A}|0\ra&=&f_V M_V\ep^\mu(h),
 \eea
where $\ep(h)$ is the polarization vector of the
vector meson. In the above definitions for the decay constants,
the experimental values of pion and rho meson decay constants are
$f_\pi\approx 131$ MeV from $\pi\to\mu\nu$ and $f_\rho\approx 220$
MeV from $\rho\to e^+e^-$.

Using Eqs.~(\ref{PPF1}) and~(\ref{fpv1}), we obtain the following expressions
for the factorized matrix elements
 \be\label{CIF1}
 X^{(B^+_c F_1,P^+)}=-if_{P}(M^2_{B_c}-M^2_{F_1})f^{B_c\to
F_1}_0(M^2_{P}),
 \ee
when $F_2$ is a pseudoscalar meson, and
 \be\label{CIF2}
 X^{(B^+_c F_1,V^+)}= 2f_{V}M_V(\epsilon\cdot
P_{B_c})f^{B_c\to F_1}_+(M^2_{V}),
 \ee
when $F_2$ is a vector meson. For the latter case, only
longitudinally polarized vector mesons are produced in the rest
frame of the decaying $B_c$ meson, i.e.,
 \be
 \epsilon\cdot P_{B_c}= \frac{M_{B_c}}{M_V}p_c,
 \ee
 where
 \be\label{k_f}
 p_c=\frac{\sqrt{[M^2_{B_c}-(M_1+M_2)^2][M^2_{B_c}-(M_1-M_2)^2]}}{2M_{B_c}},
 \ee
is the
center of mass
momentum of the final state meson($F_1$ with $M_1$ or
$F_2$ with $M_2$).

\begin{figure}
\vspace{0.8cm}
\includegraphics[width=2.5in,height=2in]{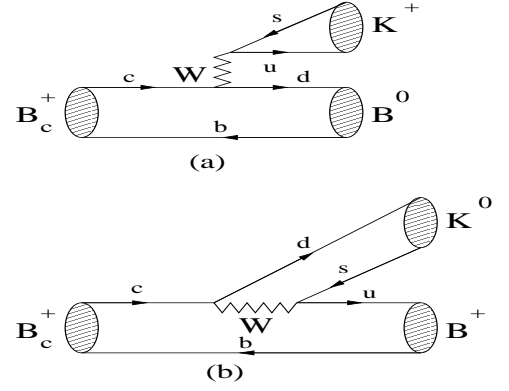}
\caption{Quark diagrams for the nonleptonic $B_c\to BK$ decays:
(a) color-favored(class I) $B^+_c\to B^{0}K^+$ and (b)
color-suppressed(class II) $B^+_c\to B^{+}K^0$ decays.}
\label{fig1n}
\end{figure}

\begin{figure}
\vspace{0.8cm}
\includegraphics[width=2.5in,height=2in]{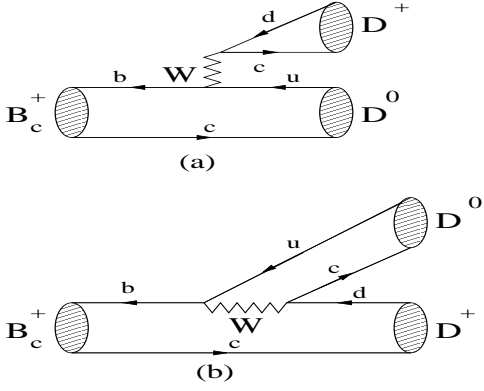}
\caption{Quark diagrams for the class III nonleptonic $B_c\to D^+D^0$ decay, which
consist of both color-favored (a) and color-suppressed (b) decays.}
\label{fig2n}
\end{figure}

The decay rate for $B_c\to F_1 + F_2$ in the rest frame of the $B_c$ meson is given by
 \be\label{BPP}
 \Gamma(B_c\to F_1F_2)=\frac{p_c}{8\pi M^2_{B_c}}
 |\la F_1F_2|{\cal H}_{\rm eff}|B^+_c\ra|^2.
 \ee
In Fig.~\ref{fig1n}, we show the example of quark diagrams for the
nonleptonic $B_c\to BK$ decays: (a) color-favored(class I)
$B^+_c\to B^{0}K^+$ and (b) color-suppressed(class II) $B^+_c\to
B^{+}K^0$ decays. We also show in Fig.~\ref{fig2n} the example of quark diagrams for the
class III transitions such as $B_c\to D^+D^0$. In the factorization approximation, these
nonleptonic decay amplitudes can be expressed as the product of one-particle
matrix elements.

\subsection{Class I decay modes}
\noindent (1) For the $b\to (u,c)(q_1\bar{q}_2)$ process,
 \be\label{cI1}
 \la D^0 M^+|{\cal H}_{\rm eff}|B^+_c\ra
=\frac{G_F}{\sqrt{2}}V_{ub}V^*_{q_1q_2}a_1 X^{(B^+_c D^0,M^+)},
 \ee
 and
 \be\label{Bcpk}
\la \eta_c M^+|{\cal H}_{\rm eff}|B^+_c\ra
=\frac{G_F}{\sqrt{2}}V_{cb}V^*_{q_1q_2}a_1 X^{(B^+_c\eta_c,M^+)},
 \ee
(2) For the $c\to(d,s)(q_1\bar{q}_2)$ process,
 \be\label{cI2}
 \la B^{0}M^+|{\cal H}_{\rm eff}|B^+_c\ra
 =\frac{G_F}{\sqrt{2}}V_{cd}V^*_{q_1q_2}a_1 X^{(B^+_c
B^{0},M^+)},
 \ee
 and
  \be
 \la B^{0}_{s}M^+|{\cal H}_{\rm eff}|B^+_c\ra
 =\frac{G_F}{\sqrt{2}}V_{cs}V^*_{q_1q_2}a_1 X^{(B^+_c
B^{0}_{s},M^+)},
 \ee
 where $M=\pi,K,\rho,K^*$ and $V_{q_1q_2}=V_{ud}$ or $V_{us}$ depending on
 whether $M=(\pi,\rho)$ or $(K,K^*)$.

\subsection{Class II decay modes}
\noindent (1) For the $b\to (d,s)(q_1\bar{q}_2)$ process,
 \be\label{CII1}
\la D^{+}M^0|{\cal H}_{\rm eff}|B^+_c\ra
=\frac{G_F}{\sqrt{2}}V_{q_2b}V^*_{q_1d}a_2 X^{(B^+_c D^{+},M^0)},
 \ee
 and
 \be\label{CII1e}
\la D^{+}_s M^0|{\cal H}_{\rm eff}|B^+_c\ra
=\frac{G_F}{\sqrt{2}}V_{q_2b}V^*_{q_1s}a_2 X^{(B^+_c
D^{+}_s,M^0}),
 \ee
(2) For the $c\to u(q_1\bar{q}_2)$ process,
 \be\label{CII2}
\la B^{+}M^0|{\cal H}_{\rm eff}|B^+_c\ra
=\frac{G_F}{\sqrt{2}}V_{cq_1}V^*_{uq_2}a_2 X^{(B^+_c B^{+},M^0)},
 \ee
and
 \bea\label{CII2e}
\la B^{+}\eta^{(\prime)}|{\cal H}_{\rm eff}|B^+_c\ra
&=&\frac{G_F}{\sqrt{2}}a_2 [ V_{cd}V^*_{ud} X^{(B^+_c
B^{+},\eta^{(\prime)}_q)} \nonumber\\
&&+ V_{cs}V^*_{us} X^{(B^+_c B^{+},\eta^{(\prime)}_s)}],
 \eea
 where $M=(\pi,\eta^{(\prime)}, \rho,\omega,\bar{D}^{(*)} )$
 for $c\to (d,s)$ induced decays and $M=(\pi,\rho,\omega,K,\bar{K},K^*,\bar{K}^{*})$
 for $c\to u$ induced decays, respectively.
 As in the case of color-favored class I
decay modes, the factorized matrix elements for color-suppressed
class II decay modes can be obtained from Eqs.~(\ref{CIF1})
and~(\ref{CIF2}) except that the decay constants for the neutral
$\pi^0,\rho^0$, and $\omega$ mesons are replaced by
$f_{P(V)}/\sqrt{2}$.

\subsection{Class III decay modes}
For the class III $B^+_c\to D^+_q M^0(q=d,s)$ transitions, the
decay amplitude is given by
 \bea\label{BcD}
\la D^+_q M^0|{\cal H}_{\rm eff}|B^+_c\ra
&=&\frac{G_F}{\sqrt{2}}V_{cq}V^*_{F}
  [a_1 X^{(B^+_cM^0,D^+_q)}
  \nonumber\\
  &&+
   a_2 X^{(B^+_cD^+_q,M^0)}],
 \eea
where $V_F=V_{cb}$ for $M=\eta_c$ and $V_{ub}$ for $M=D$. The
expressions for the factorized matrix elements can be obtained
from Eqs.~(\ref{CIF1}) and~(\ref{CIF2}).

\section{Input Parameters}

In this section we shall briefly discuss and summarize all of the
input parameters, such as the model
parameters, decay constants, and form factors for semileptonic
$B_c\to (D_{(s)},\eta_c,B_{(s)})$ decays, which are relevant to the present work.

\subsection{Brief review of LFQM}
 The key idea in our
LFQM~\cite{CJ1,CJ2} for the ground state mesons is to treat the
radial wave function as a trial function for the variational
principle to the QCD-motivated effective Hamiltonian saturating
the Fock state expansion by the constituent quark and antiquark.
The QCD-motivated effective Hamiltonian for a description of the ground
state meson mass spectra is given by
 \be\label{Ham}
H_{q\bar{q}}= H_0 + V_{q\bar{q}}= \sqrt{m^2_q+{\vec k}^2}+\sqrt{m^2_{\bar{q}}+{\vec
k}^2}+V_{q\bar{q}},
 \ee
 where
 \be\label{pot}
 V_{q\bar{q}}=V_0 + V_{\rm hyp} = a + br^n-\frac{4\al_s}{3r} +\frac{2}{3}\frac{{\bf S}_q\cdot{\bf
S}_{\bar{q}}}{m_qm_{\bar{q}}} \nabla^2V_{\rm coul}.
 \ee
In this work, we use two interaction potentials: (1) the Coulomb plus linear
confining(i.e. $n=1$) potential and (2) the Coulomb plus harmonic oscillator(HO)(i.e. $n=2$)
potential, together with the hyperfine
interaction $\la{\bf S}_q\cdot{\bf S}_{\bar{q}}\ra=1/4(-3/4)$ for
the vector (pseudoscalar) meson, which enables us to analyze the
meson mass spectra and various wave-function-related observables,
such as decay constants, electromagnetic form factors of mesons in a spacelike
region, and the weak form factors for the exclusive semileptonic
and rare decays of pseudoscalar mesons in the timelike
region~\cite{CJ1,CJ2,JC,CJK02,Choi07,Choi08}.

The momentum-space light-front wave function of the ground state
pseudoscalar and vector mesons is given by $\Psi(x_i,{\bf
k}_{i\perp},\lam_i) ={\cal R}_{\lam_1\lam_2}(x_i,{\bf k}_{i\perp})
\phi(x_i,{\bf k}_{i\perp})$, where $\phi(x_i,{\bf k}_{i\perp})$ is
the radial wave function and ${\cal R}_{\lam_1\lam_2}$ is the
covariant spin-orbit wave function. The model wave function is represented by the
Lorentz-invariant variables, $x_i=p^+_i/P^+$, ${\bf
k}_{i\perp}={\bf p}_{i\perp}-x_i{\bf P}_\perp$ and $\lam_i$, where
$P^\mu=(P^+,P^-,{\bf P}_\perp) =(P^0+P^3,(M^2+{\bf
P}^2_\perp)/P^+,{\bf P}_\perp)$ is the momentum of the meson $M$,
and $p^\mu_i$ and $\lam_i$ are the momenta and the helicities of
constituent quarks, respectively.

The covariant forms of the spin-orbit wave functions
for pseudoscalar and vector mesons are given by
 \bea\label{R00_A}
{\cal R}_{\lam_1\lam_2}^{00}
&=&\frac{-\bar{u}_{\lam_1}(p_1)\gamma_5 v_{\lam_2}(p_2)}
{\sqrt{2}\tilde{M_0}},
\nonumber\\
{\cal R}_{\lam_1\lam_2}^{1J_z}
&=&\frac{-\bar{u}_{\lam_1}(p_1)
\biggl[/\!\!\!\ep(J_z) -\frac{\ep\cdot(p_1-p_2)}{M_0 + m_1 + m_2}\biggr]
v_{\lam_2}(p_2)} {\sqrt{2}\tilde{M_0}},
\nonumber\\
 \eea
where $\tilde{M_0}=\sqrt{M^2_0-(m_1-m_2)^2}$,
$M^2_0=\sum_{i=1}^2({\bf k}^2_{i\perp}+m^2_i)/x_i$ is
the boost invariant meson mass square obtained from the free
energies of the constituents in mesons, and
$\ep^\mu(J_z)$ is the polarization vector of the vector
meson~\cite{CJ08D}.
For the radial wave function $\phi$, we use the same
Gaussian wave function for both pseudoscalar and vector mesons:
\be\label{rad}
 \phi(x_i,{\bf
k}_{i\perp})=\frac{4\pi^{3/4}}{\beta^{3/2}} \sqrt{\frac{\partial
k_z}{\partial x}} {\rm exp}(-{\vec k}^2/2\beta^2),
 \ee
 where $\beta$ is the variational parameter and
 $\sqrt{\partial k_z /\partial x}$ is the Jacobian of the variable
 transformation $\{x,{\bf k}_\perp\}\to {\vec k}=({\bf k}_\perp, k_z)$.

We apply our
variational principle to the QCD-motivated effective Hamiltonian
first to evaluate the expectation value of the central Hamiltonian
$H_0+V_0$, i.e., $\la\phi|(H_0+V_0)|\phi\ra$, with a trial
function $\phi(x_i,{\bf k}_{i\perp})$ that depends on the
variational parameter $\beta$. Once the model
parameters are fixed by minimizing the expectation value
$\la\phi|(H_0+V_0)|\phi\ra$, the mass eigenvalue of each meson is
obtained as $M_{q\bar{q}}=\la\phi|(H_0+V_{q\bar{q}})|\phi\ra$.
 Minimizing energies with
respect to $\beta$ and searching for a fit to the observed ground
state meson spectra, our central potential $V_0$ obtained from our
optimized potential parameters ($a=-0.72$ GeV, $b=0.18$ GeV$^2$,
and $\al_s=0.31$)~\cite{CJ1} for the Coulomb plus linear potential
was found to be quite comparable with the quark potential model
suggested by Scora and Isgur~\cite{SI}, where they obtained
$a=-0.81$ GeV, $b=0.18$ GeV$^2$, and $\al_s=0.3\sim 0.6$ for the
Coulomb plus linear confining potential. A more detailed procedure
for determining the model parameters of light- and heavy-quark
sectors can be found in our previous works~\cite{CJ1,CJ2}.

\begin{table*}[t]
\caption{The constituent quark mass[GeV] and the Gaussian
parameters $\beta$[GeV] for the linear and HO potentials obtained
by the variational principle. $q=u$ and $d$.}\label{t2n}
\begin{tabular}{ccccccccccccccc} \hline\hline
Model & $m_q$ & $m_s$ & $m_c$ & $m_b$ & $\beta_{qq}$ &
$\beta_{qs}$  & $\beta_{ss}$ & $\beta_{qc}$ & $\beta_{sc}$ &
$\beta_{cc}$ & $\beta_{qb}$ & $\beta_{sb}$
& $\beta_{cb}$ & $\beta_{bb}$ \\
\hline Linear & 0.22 & 0.45 & 1.8 & 5.2 & 0.3659 & 0.3886 & 0.4128
&
0.4679 & 0.5016 & 0.6509 & 0.5266 & 0.5712 & 0.8068 & 1.1452\\
\hline HO & 0.25 & 0.48 & 1.8 & 5.2 & 0.3194 & 0.3419 & 0.3681 &
0.4216 & 0.4686 & 0.6998 & 0.4960 & 0.5740 & 1.0350 &1.8025\\
\hline\hline
\end{tabular}
\end{table*}
Our model parameters $(m_q,\beta_{q\bar{q}})$  obtained from the linear and
HO potential models are summarized in Table~\ref{t2n}. The
predictions of the ground state meson mass spectra including
bottom-charmed mesons can be found in our
recent work, Ref.~\cite{CJBc}.

\subsection{Form factors for semileptonic $B_c\to P$ decays}

For the nonleptonic two-body $B_c$ decays,
we use the $q^2$ dependent form factors $f_+(q^2)$ and $f_0(q^2)$
for the $B_c\to(D,D_s,\eta_c,B,B_s)$ decays as input parameters.

Within the framework of LF quantization, while the form factor
$f_+(q^2)$ can be obtained only from the valence contribution in the
$q^{+}= 0$ frame with the ``$+$" component of the currents
without encountering the zero-mode contribution~\cite{Zero}, the form
factor $f_-(q^2)$ [or equivalently $f_0(q^2)$] receives the higher Fock state
contribution (i.e., the zero-mode in the $q^+=0$ frame or the nonvalence contribution
in the $q^+> 0$ frame). In order to calculate $f_-(q^2)$,
we developed in~\cite{JC,CJK02} an effective treatment of handling
the higher Fock state (or nonvalence) contribution to $f_-(q^2)$ in
the purely longitudinal $q^+>0$ frame (i.e., $q^2=q^+q^->0$) based
on the
Bethe-Salpeter(BS)
formalism.
In our recent LFQM analysis~\cite{CJBc} of the semileptonic
$B_c\to(D,\eta_c,B,B_s)\ell\nu_\ell$ decays, we utilized our effective method~\cite{JC}
to express the zero-mode contribution as a convolution of zero-mode operator
with the initial and final state LF wave functions.
In this way, we obtained the form factor $f_-(q^2)$ in the $q^+=0$ frame
using the perpendicular components of the currents and discussed
the LF covariance of $f_-(q^2)$ in the valence region by analyzing
the covariant BS model and the LF covariant analysis
described by Jaus~\cite{Jaus99}.

The LF covariant form factors $f_+(q^2)$ and $f_-(q^2)$ for
$B_c(q_1\bar{q})\to P(q_2\bar{q})$ transitions
obtained from the $q^+=0$ frame are given by (see
~\cite{CJBc} for more detailed derivations)
\begin{widetext}
 \bea\label{fpp}
 f_{+}(q^2) &=& \int^{1}_{0}dx\int
\frac{d^{2}{\bf k}_{\perp}}{16\pi^3}
\frac{\phi_{1}(x,{\bf k}_{\perp})}{\sqrt{ {\cal A}_{1}^{2}
 + {\bf k}^{2}_{\perp}}}
\frac{\phi_{2}(x,{\bf k}'_{\perp})}{\sqrt{ {\cal A}_{2}^{2}
+ {\bf k}^{\prime 2}_{\perp}}}
 ( {\cal A}_{1}{\cal A}_{2}+{\bf k}_{\perp}\cdot{\bf k'}_{\perp} ),
     \nonumber\\
 f_-(q^2) &=& \int^1_0 (1-x) dx
 \int \frac{ d^2{\bf k}_\perp } { 16\pi^3 }
  \frac{ \phi_1 (x, {\bf k}_\perp) } {\sqrt{ {\cal A}^2_1 + {\bf k}^2_\perp }}
  \frac{ \phi_2 (x, {\bf k'}_\perp) } {\sqrt{ {\cal A}^2_2 + {\bf k}^{\prime 2}_\perp }}
   \biggl\{ -x(1-x) M^2_1 - {\bf k}^2_\perp - m_1m_{\bar{q}} + (m_2 - m_{\bar{q}}){\cal A}_1
 \nonumber\\
 && + 2\frac{q\cdot P}{q^2} \biggl[ {\bf k}^2_\perp
 + 2\frac{ ( {\bf k}_\perp \cdot {\bf q}_\perp)^2 } {q^2} \biggr]
 + 2 \frac{ ( {\bf k}_\perp \cdot {\bf q}_\perp)^2 } {q^2}
  + \frac{ {\bf k}_\perp \cdot {\bf q}_\perp } {q^2}  [ M^2_2 - (1-x) (q^2 + q\cdot P) + 2 x M^2_0
  \nonumber\\
  && - (1 - 2x) M^2_1 - 2(m_1 - m_{\bar{q}}) (m_1 + m_2) ] \biggr\},
 \eea
\end{widetext}
where ${\bf k'}_\perp={\bf k}_\perp + (1-x){\bf q}_\perp$,
${\cal A}_{i}= (1-x) m_{i} + x m_{\bar{q}}$ ($i=1,2$), and $q\cdot P=M^2_1-M^2_2$ with
$M_1$ and $M_2$ being the physical masses of the initial and final
state
mesons, respectively.
We should note that the LF covariant form factor $f_-(q^2)$ in Eq.~(\ref{fpp}) is the sum
of the valence contribution $f^{\rm val}_-(q^2)$ and the zero-mode contribution $f^{\rm Z.M.}_-(q^2)$.

\begin{figure}
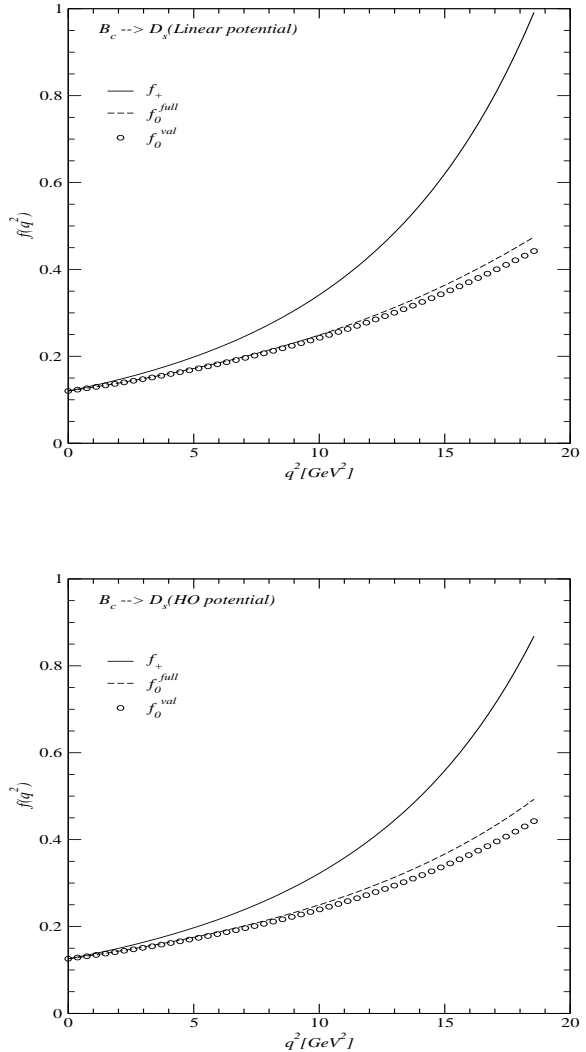

\vspace{0.5cm}
\includegraphics[width=3in,height=2.5in]{Fig3a.eps}\\
\vspace{1.2cm}
\includegraphics[width=3in,height=2.5in]{Fig3b.eps}
\caption{The weak form factors $f_+(q^2)$ (solid
line) and $f_0(q^2)$ (dashed line) for the $B_c\to D_s$ transition obtained
from the linear (upper panel) and HO (lower panel) potential parameters. The circles
represent the valence contributions $f^{\rm val}_0(q^2)$ to $f_0(q^2)$.
} \label{fig3n}
\end{figure}
For the analysis of the nonleptonic $B_c\to D_s F$ decays where
$F$ is the vector or pseudoscalar meson being factored out, we
show in Fig.~\ref{fig3n} the $q^2$-dependence of the weak form
factors $f_+(q^2)$ (solid line) and $f_0(q^2)$ (dashed line) for the
$B_c\to D_s$ transition obtained from the linear (upper panel) and
HO (lower panel) potential parameters. The circles
represent the valence contribution $f^{\rm val}_0(q^2)$ to $f_0(q^2)$.
That is, the difference between $f_0(q^2)$ and $f^{\rm val}_0(q^2)$ represents
the zero-mode contribution
to $f_0(q^2)$.
We obtain $f_+(0)=f_0(0)=0.120$ [0.126] at $q^2=0$ for the
linear [HO] potential model. The form factors at the zero-recoil point (i.e.
$q^2=q^2_{\rm max})$ are obtained as $f_+(q^2_{\rm
max})=0.992$ [0.868] and $f_0(q^2_{\rm
max})=0.475$ [0.493] for the linear [HO] potential model.
On the other hand, the valence contribution to $f_0(q^2)$ at
the zero-recoil point is obtained as
$f^{\rm val}_0(q^2_{\rm max})=0.442$ [0.443] for the linear [HO] potential model.

In Table~\ref{t3n}, we show the decay form factors $f_+(0)=f_0(0)$
at $q^2=0$ for the semileptonic $B_c\to(D,\eta_c,B,B_s)$ decays
obtained from~\cite{CJBc} and the rare $B_c\to D_s$ decay obtained in the
present work (i.e. Fig.~\ref{fig3n}) and compare them to other theoretical model
predictions.

\begin{table*}[t]
\caption{Form factors $f_+(0)=f_0(0)$ at $q^2=0$  for $B_c\to
(D_{(s)},\eta_c, B, B_s)$  transitions.}\label{t3n}
\begin{tabular}{ccccccccc} \hline\hline
 &  Linear[HO] & ~~\cite{EFG03D,EFG03E} & ~~\cite{IKS01} &
~\cite{NW}& ~\cite{HNV} & ~\cite{AKN} & ~\cite{WSL} & ~\cite{DSV} \\
\hline $f_{+}^{B_c\to D}(0)$ & 0.086[0.079] & 0.14 & 0.69& 0.1446 & - & 0.089 & 0.16
& $0.08\pm 0.02$ \\
\hline $f_{+}^{B_c\to D_s}(0)$ & 0.120[0.126] &-&-&-&-&-&0.28& $0.15\pm 0.02$\\
 \hline $F_+^{B_c\to\eta_c}(0)$ &
0.482[0.546] & 0.47 &
0.76& 0.5359 & 0.49 & 0.622 & 0.61 & $0.58$\\
 \hline $f_+^{B_c\to B}(0)$ & 0.467[0.426] & 0.39 &
0.58& 0.4504 & 0.39 & 0.362 & 0.63 & $0.41\pm 0.04$\\
 \hline $f_+^{B_c\to B_s}(0)$ & 0.573[0.571] & 0.50 &
0.61& 0.5917 & 0.58 & 0.564 & 0.73 & $0.55\pm 0.03$\\
\hline\hline
\end{tabular}
\end{table*}

\subsection{Weak decay constants of $\eta$ and $\eta'$}

In this work, we shall also consider the nonleptonic decays of
$B_c$ mesons to isoscalar states such as $\omega$ and
$(\eta,\eta')$. Isoscalar states with the same $J^{PC}$ will mix,
but mixing between the two light-quark isoscalar mesons, and the
much heavier charmonium or bottomonium states is generally
assumed to be negligible. Since the vector mixing angle is known
to be very close to ideal mixing, we assume ideal mixing between
$\omega$ and $\phi$ mesons, i.e.,
$\omega=(u\bar{u}+d\bar{d})/\sqrt{2}$ and $\phi=s\bar{s}$.
However, the octet-singlet mixing angle $\theta$ of $\eta$ and $\eta'$
is known to be in the range of $-10^{\rm o}$ to $-23^{\rm o}$.
The physical $\eta$ and $\eta'$ are the mixtures of the flavor $SU(3)$ octet
$\eta_8$ and singlet $\eta_0$ states:
 \be\label{eet}
 \left( \begin{array}{cc}
 \eta\\
 \eta'
 \end{array}\,\right)
 =U(\theta) \left( \begin{array}{c}
 \eta_8\\
 \eta_0
 \end{array}\,\right),
 \ee
 where
 \be\label{uni}
 U(\theta)=\left( \begin{array}{cc}
 \cos\theta\;\; -\sin\theta\\
 \sin\theta\;\;\;\;\;\cos\theta
 \end{array}\,\right),
 \ee
 and
$\eta_8=(u\bar{u}+d\bar{d}-2s\bar{s})/\sqrt{6}$ and
$\eta_0=(u\bar{u}+d\bar{d} + s\bar{s})/\sqrt{3}$. Analogously,
in terms of the quark-flavor basis $\eta_q=(u\bar{u}+d\bar{d})/\sqrt{2}$ and
 $\eta_s=s\bar{s}$, one obtains~\cite{FKS}
  \be\label{eea}
 \left( \begin{array}{cc}
 \eta\\
 \eta'
 \end{array}\,\right)
 =U(\phi)\left( \begin{array}{c}
 \eta_q\\
 \eta_s
 \end{array}\,\right).
 \ee
 The two schemes are equivalent to each
 other by $\phi=\theta+ \arctan\sqrt{2}$ when ${\rm SU}_f(3)$ symmetry is perfect.
 However, when one takes into account the ${\rm SU}_f(3)$ breaking effect,
 this relationship is not maintained but given by the following Fock decompositions
 of the octet-singlet basis states~\cite{FKS}:
 \bea\label{osb}
 |\eta_8\ra &=& \frac{\Psi_q+2\Psi_s}{3}
 \frac{|u\bar{u}+d\bar{d}-2s\bar{s}\ra}{\sqrt{6}}
 \nonumber\\
 &&+\frac{\sqrt{2}(\Psi_q-\Psi_s)}{3}
  \frac{|u\bar{u}+d\bar{d}+s\bar{s}\ra}{\sqrt{3}},
  \nonumber\\
  |\eta_0\ra &=& \frac{\sqrt{2}(\Psi_q-\Psi_s)}{3}
 \frac{|u\bar{u}+d\bar{d}-2s\bar{s}\ra}{\sqrt{6}}
 \nonumber\\
 &&+\frac{2\Psi_q+\Psi_s}{3}
  \frac{|u\bar{u}+d\bar{d}+s\bar{s}\ra}{\sqrt{3}},
  \eea
where $\Psi_i$ denote LF wave functions of the corresponding parton states.
Only in the ${\rm SU}_f(3)$ symmetry limit, i.e., $\Psi_q=\Psi_s$,  would one find
pure octet and singlet states in Eq.~(\ref{osb}).
Although it was
frequently assumed that the decay constants follow the same
pattern of state mixing, the mixing properties of
the decay constants will generally be different from the mixing
properties of the meson state since the decay constants only probe
the short-distance properties of the valence Fock states while the
state mixing refers to the mixing of the overall wave
function~\cite{FKS}.

Using the decay constants of Eq.~(\ref{fpv1}) defined in the quark-flavor basis,
the two basic decay constants $f_q$ and
$f_s$ arising from $\eta_q$ and $\eta_s$ are obtained as
\be\label{fqs}
f_{q(s)}=2\sqrt{6}\int^1_0 dx\int\frac{d^2{\bf k}_\perp}{16\pi^3}
\Psi_{q(s)},
\ee
and simply follow the pattern of state mixing due to the 
Okubo-Zweig-Iizuka(OZI) rule~\cite{OZI}, i.e,
 \be\label{fqfs}
 \left( \begin{array}{cc}
 f^q_\eta  \;\;\;\;\; f^s_\eta\\
 f^q_{\eta'} \;\;\;\;\; f^s_{\eta'}
 \end{array}\,\right)
 = U(\phi)\left( \begin{array}{cc}
 f_q\;\; 0\\
 0\;\;\;f_s
 \end{array}\,\right).
 \ee
The OZI rule implies that the difference between the two mixing angles
$\phi_q$ and $\phi_s$ vanishes (i.e., $\phi_q=\phi_s=\phi$ in Eq.~(\ref{fqfs})
to leading order in the $1/N_c$ expansion.
On the other hand, the decay constants in the
octet-singlet basis are parametrized as~\cite{FKS,Leut98}
 \be\label{feet}
 \left( \begin{array}{cc}
 f^8_\eta  \;\;\;\;\; f^0_\eta\\
 f^8_{\eta'} \;\;\;\;\; f^0_{\eta'}
 \end{array}\,\right)
 =\left( \begin{array}{cc}
 \cos\theta_8\;\; -\sin\theta_0\\
 \sin\theta_8\;\;\;\;\;\cos\theta_0
 \end{array}\,\right)\left( \begin{array}{cc}
 f_8\;\; 0\\
 0\;\;\;f_0
 \end{array}\,\right),
 \ee
 where $\theta_8$ and $\theta_0$ turn out to differ
 considerably and become equal only in the ${\rm SU}_f(3)$ symmetry limit.

 By using the correlation between the quark-flavor mixing scheme and the
 octet-singlet scheme~\cite{FKS,OZI}, one obtains
 \bea\label{4p}
 f_8&=&\sqrt{\frac{f^2_q+2f^2_s}{3}},\;\theta_8=\phi - \arctan(\sqrt{2}f_s/f_q),
 \nonumber\\
 f_0&=&\sqrt{\frac{2f^2_q+f^2_s}{3}},\;\theta_0= \phi -
 \arctan(\sqrt{2}f_q/f_s).
 \eea
In our previous work~\cite{CJ1}, we obtained  the $\eta-\eta'$
mixing angle $\theta\simeq -19^{\rm o}$ for both linear and HO
potential models by fitting the physical masses of $\eta$ and
$\eta'$. This corresponds to the mixing angle $\phi=35.7^{\rm o}$
in the quark-flavor basis. We applied this mixing angle to predict the
decay widths for $\eta(\eta')\to\gamma\gamma$ using the axial
anomaly plus partial conservation of the axial vector
current (PCAC) relations~\cite{PCAC} and obtained $f_8/f_\pi=1.32\;
(1.25)$ and $f_0/f_\pi=1.16 \;(1.13)$ for the linear (HO)
potential model~\cite{CJ1}. From the decay constants of octet and
singlet mesons together with Eq.~(\ref{4p}), we now obtain the
four parameters ($f_q, f_s, \theta_8$,$\theta_0$) as follows:
$f_q/f_\pi= 0.97\; (1.00)$, $f_s/f_\pi=1.46\; (1.36)$,
$\theta_8=-29.2^{\rm o}\; (-26.8^{\rm o})$, and
$\theta_0=-7.3^{\rm o}\; (-10.6^{\rm o})$ for the linear (HO)
potential model, respectively. Given this background, we finally
obtain the decay constants related to the $\eta$ and $\eta'$
mesons as follows
 \begin{eqnarray}\label{fqs_v}
 &&f^q_\eta = 103.2\; (106.4)\;{\rm MeV}, \;\;
 f^s_\eta = -116.6\; (-104.0)\; {\rm MeV},
 \nonumber\\
 &&f^q_{\eta'}= 74.2\; (76.4)\;{\rm MeV},\;\;\;\;\;
 f^s_{\eta'}=155.3\;
 (144.7)\;{\rm MeV}.
 \nonumber\\
 \end{eqnarray}
Our results for the mixing parameters of $\eta$ and $\eta'$ are
consistent with those obtained from Feldmann {\em et
al.}~\cite{FKS}, namely, $f_8/f_\pi=1.26$, $f_0/f_\pi=1.17$,
$f_q/f_\pi=(1.07\pm 0.02)$, $f_s/f_\pi=(1.34\pm 0.06)$,
$\phi=(39.3^{\rm o}\pm 1.0^{\rm o})$, $\theta_8=-21.2^{\rm o}$,
$\theta_0=-9.2^{\rm o}$,
 $f^q_\eta = (108.5\pm 2.0)$ MeV,
 $f^s_\eta = -(111.2\pm 5.0)$ MeV,
 $f^q_{\eta'}= (88.8\pm 1.7)$ MeV, and
 $f^s_{\eta'}=(135.8\pm 6.1)$ MeV.

For the measured values of meson decay constants, we use the
central values extracted from the experimental
measurements~\cite{Data08,Artuso,Cleo09}. However, for the
unmeasured decay constants, we use the average values obtained
from our linear and HO model predictions in~\cite{Choi07} in addition to
the present work. The values for the decay constants used in this
work are compiled in Table~\ref{t4n}.
\begin{table*}
\caption{Meson decay constants(in unit of MeV) used in this work.
}\label{t4n}
\begin{tabular}{ccccccccccccc} \hline\hline
$f_\pi$ & $f_K$ & $f_\rho$ & $f_\omega$ & $f_{K^*}$ & $f^q_\eta$ &
$f^s_\eta$ & $f^q_{\eta'}$ & $f^s_{\eta'}$ &$f_D$ & $f_{D^*}$ &
$f_{\eta_c}$ &
$f_{D_s}$\\
\hline
 131~\cite{Data08} & 159.8~\cite{Data08} & 220~\cite{Data08} &
 195~\cite{Data08}&
 217~\cite{Data08} & 104.8 & -110.3&
75.3 & 150.0 & 222.6~\cite{Artuso} & 241~\cite{Choi07} &
340~\cite{Choi07}
& 259.5~\cite{Cleo09}\\
 \hline\hline
\end{tabular}
\end{table*}
Since the decay constant $f_{\eta_c}$ extracted
from CLEO Collaboration~\cite{Edwards} has a
large error bar, i.e., $f^{\rm CLEO}_{\eta_c}=335\pm 75$ MeV, we
instead take the average value $f_{\eta_c}=340$ MeV of our LFQM
predictions~\cite{Choi07}, i.e., $f^{\rm lin}_{\eta_c}=326$ MeV
and $f^{\rm HO}_{\eta_c}=354$ MeV.

\begin{table*}[t]
\caption{Exclusive nonleptonic decay widths $\Gamma$ (in
$10^{-15}$ GeV) of the $B_c$ meson for the general values of the
Wilson coefficients $a_1$ and $a_2$.}\label{t5}
\begin{tabular}{clcccc} \hline\hline
Class & Mode & Lin [HO] & ~~\cite{EFG03D,EFG03E}& ~~\cite{HNV}& ~~\cite{AKN}\\
\hline
 &$B^+_c\to D^0\pi^+$ & $4.7[4.7]$$(\times 10^{-4})a^2_1$ & - & - & -\\
 &$B^+_c\to D^0\rho^+$ & 1.4[1.2]$(\times 10^{-3})a^2_1$ & - & - & -\\
 &$B^+_c\to D^0K^+$ & $3.9[3.4]$$(\times 10^{-5})a^2_1$ & - & - & -\\
&$B^+_c\to D^0K^{*+}$ & $7.5[6.4]$$(\times 10^{-5})a^2_1$& - & - & -\\
&$B^+_c\to\eta_c\pi^+$ & 0.997[1.280]$a^2_1$ & 0.93$a^2_1$ & 1.02$a^2_1$& 1.47$a^2_1$\\
&$B^+_c\to\eta_c \rho^+$ & 2.827[3.563]$a^2_1$& 2.3$a^2_1$ & 2.60$a^2_1$& 3.35$a^2_1$\\
&$B^+_c\to\eta_c K^+$ & 0.081[0.103]$a^2_1$ & 0.073$a^2_1$ & 0.082$a^2_1$& 0.15$a^2_1$\\
I &$B^+_c\to\eta_c K^{*+}$ & 0.147[0.184]$a^2_1$ & 0.12$a^2_1$ & 0.15$a^2_1$& 0.24$a^2_1$\\
&$B^+_c\to B^0\pi^+$ & 1.557[1.296]$a^2_1$ & 1.0$a^2_1$ & 1.10$a^2_1$& 1.51$a^2_1$\\
&$B^+_c\to B^0\rho^+$ & 1.936[1.505]$a^2_1$ & 1.3$a^2_1$ & 1.41$a^2_1$& 1.93$a^2_1$\\
&$B^+_c\to B^0 K^+$ & 0.126[0.104]$a^2_1$ & 0.09$a^2_1$ & 0.098$a^2_1$& -\\
&$B^+_c\to B^0 K^{*+}$ & 0.042[0.032]$a^2_1$ & 0.04$a^2_1$ & 0.038$a^2_1$& -\\
&$B^+_c\to B^0_s\pi^+$ & 36.97[36.71]$a^2_1$& 25$a^2_1$ & 34.7$a^2_1$& 34.78$a^2_1$\\
&$B^+_c\to B^0_s\rho^+$ & 25.43[23.22]$a^2_1$&  14$a^2_1$ & 23.1$a^2_1$& 23.61$a^2_1$\\
&$B^+_c\to B^0_s K^+$ & 2.853[2.816]$a^2_1$ & 2.1$a^2_1$ & 2.87$a^2_1$& -\\
&$B^+_c\to B^0_s K^{*+}$ & 0.069[0.061]$a^2_1$ & 0.03$a^2_1$ & 0.13$a^2_1$& -\\
\hline
&$B^+_c\to D^+\pi^0$ & $2.4[2.0](\times 10^{-4})a^2_2$ & - & -& -\\
&$B^+_c\to D^+\rho^0$ & $7.0[6.0](\times 10^{-4})a^2_2$ & - & -& -\\
&$B^+_c\to D^+\omega$ & $5.5[4.7](\times 10^{-4})a^2_2$ & - & -& -\\
&$B^+_c\to D^+\eta$ & $3.1[2.7](\times 10^{-4})a^2_2$ & - & -& -\\
&$B^+_c\to D^+\eta'$ & $1.7[1.5](\times 10^{-5})a^2_2$ & - & -& -\\
&$B^+_c\to D^+\bar{D}^0$ & $0.219[0.185]a^2_2$& - & -& -\\
&$B^+_c\to D^+\bar{D}^{*0}$ & $0.261[0.212]a^2_2$ & - & -& -\\
&$B^+_c\to D^+_s\pi^0$ & $2.4[2.6](\times 10^{-5})a^2_2$ & - & -& -\\
&$B^+_c\to D^+_s\rho^0$ & $7.1[7.6](\times 10^{-5})a^2_2$ & - & -& -\\
&$B^+_c\to D^+_s\omega$ & $5.6[6.0](\times 10^{-5})a^2_2$ & - & -& -\\
&$B^+_c\to D^+_s\eta$ & $3.2[3.4](\times 10^{-5})a^2_2$ & - & -& -\\
&$B^+_c\to D^+_s\eta'$ & $1.7[1.9](\times 10^{-5})a^2_2$ & - & -& -\\
&$B^+_c\to D^+_s\bar{D}^{0}$ & $0.0216[0.0227]a^2_2$ & - & -& -\\
&$B^+_c\to D^+_s\bar{D}^{*0}$ & $0.0248[0.0250]a^2_2$ & - & -& -\\
II&$B^+_c\to B^+\pi^0$ & 0.779[0.648]$a^2_2$ & 0.5$a^2_2$ & 0.54$a^2_2$& 1.03$a^2_2$\\
&$B^+_c\to B^+ \rho^0$ & 0.967[0.752]$a^2_2$ & 0.7$a^2_2$ & 0.71$a^2_2$& 1.28$a^2_2$\\
&$B^+_c\to B^+ \omega$ & 0.721[0.558]$a^2_2$ & - & -& -\\
&$B^+_c\to B^+\eta$ & 3.99[3.30]$a^2_2$ & - & -& -\\
&$B^+_c\to B^+\eta'$ & 0.054[0.045]$a^2_2$ & - & -& -\\
&$B^+_c\to B^+ K^0$ & 0.125[0.104]$a^2_2$ & - & -& -\\
&$B^+_c\to B^+ \bar{K}^0$ & 47.85[39.66]$a^2_2$ & 34$a^2_2$ & 35.3$a^2_2$& -\\
&$B^+_c\to B^+ K^{*0}$ & 0.040[0.030]$a^2_2$ & -& -& -\\
&$B^+_c\to B^+ \bar{K}^{*0}$ & 15.36[11.38]$a^2_2$ & 13$a^2_2$& 13.1$a^2_2$& -\\
\hline
&$B^+_c\to D^+D^0$ & $(0.011 a_1 + 0.011 a_2)^2[(0.0097 a_1 + 0.0097 a_2)^2]$ & - & -& -\\
III&$B^+_c\to D^+_sD^0$ & $(0.058 a_1 + 0.064 a_2)^2[(0.052 a_1 + 0.066 a_2)^2]$ & - & -& -\\
&$B^+_c\to \eta_c D^+$ & $(0.428 a_1 + 0.226 a_2)^2[(0.482 a_1 + 0.208 a_2)^2]$ & - & $(0.438 a_1 + 0.236 a_2)^2$& $(0.47 a_1 + 0.73 a_2)^2$\\
&$B^+_c\to \eta_c D^+_s$ & $(2.27 a_1 + 1.32 a_2)^2[(2.47 a_1 + 1.34 a_2)^2]$ & - & $(2.54 a_1 + 1.93 a_2)^2$& $(2.59 a_1 + 3.40 a_2)^2$\\
 \hline\hline
\end{tabular}
\end{table*}
\begin{table*}[t]
\caption{Branching ratios (in $\%$) of the exclusive non-leptonic
$B_c$ decays at the fixed choice of Wilson coefficients:
$a^c_1\;(a^b_1)=1.20\; (1.14)$ and $a^c_2\;(a^b_2)=-0.317\;(-0.20)$
relevant for the nonleptonic decays of the $c\;(\bar{b})$ quark.
For the lifetime of the $B_c$ we take $\tau(B_c)=0.46$
ps.}\label{t6}
\begin{tabular}{clccccccccc} \hline\hline
Class & Mode & Lin [HO] & ~\cite{EFG03D,EFG03E}& ~\cite{HNV}&
~\cite{AKN}& ~\cite{IKS06} & ~\cite{Gouz} &~\cite{CD00} &~\cite{SYDM}&~\cite{CC94}\\
\hline
 &$B^+_c\to D^0\pi^+$ & $4.3[4.3]$$(\times 10^{-5})$ & - & - & -&-&-&-&-&-\\
 &$B^+_c\to D^0\rho^+$ & 1.3[1.1]$(\times 10^{-4})$ & - & - & -&-&-&-&-&-\\
 &$B^+_c\to D^0K^+$ & $3.5[3.1]$$(\times 10^{-6})$ & - & - & -&-&-&-&-&-\\
&$B^+_c\to D^0K^{*+}$ & $6.8[5.8]$$(\times 10^{-6})$& - & - & -&-&-&-&-&-\\
&$B^+_c\to\eta_c\pi^+$ & 0.091[0.116] & 0.085 & 0.094& 0.13 & 0.19 & 0.20&0.025&-&0.18\\
&$B^+_c\to\eta_c \rho^+$ & 0.257[0.324]& 0.21 & 0.24& 0.30 & 0.45 &0.42&0.067&-&0.49\\
&$B^+_c\to\eta_c K^+$ & 0.0074[0.0094] & 0.0075 & 0.0075& 0.013&0.015 &0.013&0.002&-&0.014\\
I &$B^+_c\to\eta_c K^{*+}$ & 0.013[0.017] & 0.011 & 0.013& 0.021 & 0.025 &0.020&0.004&-&0.025\\
&$B^+_c\to B^0\pi^+$ & 0.157[0.131] & 0.10 & 0.11& 0.15 & 0.20 &1.06&0.19 & 0.373 &0.32\\
&$B^+_c\to B^0\rho^+$ & 0.195[0.152] & 0.13 & 0.14& 0.19 & 0.20 &0.96&0.15 & 0.527 &0.59\\
&$B^+_c\to B^0 K^+$ & 0.013[0.011] & 0.009 & 0.010& - & 0.015 &0.07&0.014 & 0.027 &0.025\\
&$B^+_c\to B^0 K^{*+}$ & 0.0042[0.0032] & 0.004 & 0.0039& -&0.0048&0.015&0.003&0.023 &0.018\\
&$B^+_c\to B^0_s\pi^+$ & 3.723[3.697]& 2.52 & 3.51& 3.42 & 3.9&16.4&3.01 & 5.309 & 5.75\\
&$B^+_c\to B^0_s\rho^+$ & 2.561[2.338]&  1.41 & 2.34 & 2.33 & 2.3&7.2&1.34 &6.265 & 4.41\\
&$B^+_c\to B^0_s K^+$ & 0.287[0.284] & 0.21 & 0.29& - & 0.29 &1.06&0.21 & 0.367 & 0.41\\
&$B^+_c\to B^0_s K^{*+}$ & 0.0069[0.0061] & 0.003 & 0.013& -& 0.011&-&0.0043 &0.165 & -\\
\hline
&$B^+_c\to D^+\pi^0$ & $6.7[5.6](\times 10^{-7})$ & - & -& - & -&-&-&-&-\\
&$B^+_c\to D^+\rho^0$ & $2.0[1.7](\times 10^{-6})$ & - & -& -& -&-&-&-&-\\
&$B^+_c\to D^+\omega$ & $1.5[1.3](\times 10^{-6})$ & - & -& - &-&-&-&-&-\\
&$B^+_c\to D^+\eta$ & $8.7[7.6](\times 10^{-7})$ & - & -& - & -&-&-&-&-\\
&$B^+_c\to D^+\eta'$ & $4.8[4.2](\times 10^{-8})$ & - & -& - & -&-&-&-&-\\
&$B^+_c\to D^+\bar{D}^0$ & $6.1[5.2](\times 10^{-4})$& - & -&
-&$3.3\times 10^{-3}$
   & $5.3\times 10^{-3}$& $4.1\times 10^{-4}$&-& $1.8\times 10^{-3}$\\
&$B^+_c\to D^+\bar{D}^{*0}$ & $7.3[5.9](\times 10^{-4})$ & - & -&
-&$3.8\times 10^{-3}$ & $7.5\times 10^{-3}$ & $3.6\times 10^{-4}$&-& $1.9\times 10^{-3}$\\
&$B^+_c\to D^+_s\pi^0$ & $6.7[7.3](\times 10^{-8})$ & - & -& - & -&-&-&-&-\\
&$B^+_c\to D^+_s\rho^0$ & $2.0[2.1](\times 10^{-7})$ & - & -& - & -&-&-&-&-\\
&$B^+_c\to D^+_s\omega$ & $1.6[1.7](\times 10^{-7})$ & - & -& - & -&-&-&-&-\\
&$B^+_c\to D^+_s\eta$ & $9.0[9.5](\times 10^{-8})$ & - & -& - & -&-&-&-&-\\
&$B^+_c\to D^+_s\eta'$ & $4.8[5.3](\times 10^{-8})$ & - & -& - & -&-&-&-&-\\
&$B^+_c\to D^+_s\bar{D}^{0}$ & $6.0[6.3](\times 10^{-5})$ & - & -&
-&$2.1\times 10^{-4}$ & $4.8\times 10^{-4}$ & $2.7\times 10^{-5}$&- & $9.3\times 10^{-5}$\\
&$B^+_c\to D^+_s\bar{D}^{*0}$ & $6.9[7.0](\times 10^{-5})$ & - &
-& -&$2.4\times 10^{-4}$
   & $7.1\times 10^{-4}$ & $2.5\times 10^{-5}$&- & $9.7\times 10^{-5}$\\
II&$B^+_c\to B^+\pi^0$ & 0.0055[0.0046] & 0.004 & 0.0038& 0.007
&0.007 &0.037&-
          & $4.6\times 10^{-5}$ &0.011\\
 &$B^+_c\to B^+ \rho^0$ & 0.0068[0.0053] & 0.005 & 0.0050& 0.009&0.0071&0.034&-
   & $6.5\times 10^{-5}$&0.020\\
&$B^+_c\to B^+ \omega$ & 0.0051[0.0039] & - & -& -&-&-&- & $5.8\times 10^{-5}$ &-\\
&$B^+_c\to B^+\eta$ & 0.028[0.023] & - & -& -&-&-&- & $1.6\times 10^{-4}$ &-\\
&$B^+_c\to B^+\eta'$ & $3.8[3.2](\times 10^{-4})$ & - & -& -&-&-&- & $8.9\times 10^{-6}$ &-\\
&$B^+_c\to B^+ K^0$ & $8.8[7.3](\times 10^{-4})$ & - & -& -&-&-&- & $6.5\times 10^{-6}$&-\\
&$B^+_c\to B^+ \bar{K}^0$ & 0.336[0.279] & 0.24 & 0.25& -&0.38 &1.98&-& 0.0022 &0.66\\
&$B^+_c\to B^+ K^{*0}$ & $2.8[2.1](\times 10^{-4})$ & -& -& -&-&-&-& $5.5\times 10^{-6}$&-\\
&$B^+_c\to B^+ \bar{K}^{*0}$ & 0.108[0.080] & 0.09& 0.093& -&0.11&0.43&-& 0.0018 & 0.47\\
\hline &$B^+_c\to D^+D^0$ & $7.5[5.8](\times 10^{-6})$ & - & -&
-&$3.1\times 10^{-5}$
     &$3.2\times 10^{-5}$&-&-&-\\
III&$B^+_c\to D^+_sD^0$ & $2.0[1.5](\times 10^{-4})$ & - & -&
-&$7.4\times 10^{-4}$
    &$6.6\times 10^{-4}$&-&-&-\\
&$B^+_c\to \eta_c D^+$ & $0.014[0.018]$ & - & 0.014& 0.010&0.019 &0.032&0.0055&-&0.0012\\
&$B^+_c\to \eta_c D^+_s$ & $0.378[0.454]$ & - & 0.44& 0.35&0.44 &0.86 &0.51&-&0.056\\
 \hline\hline
\end{tabular}
\end{table*}

\section{Numerical Results}

In our numerical calculations of exclusive $B_c$ decays, we use
two sets of model parameters ($m,\beta$) for the linear and HO
confining potentials given in Table~\ref{t2n} to compute
the weak form factors for
semileptonic  $B_c\to (D_{(s)},\eta_c,B_{(s)})$ decays and the
unmeasured decay constants as given in Tables~\ref{t3n}
and~\ref{t4n}, respectively. Using them together with the CKM
matrix elements given by Table~\ref{t1n} we finally predict the
branching ratios which are given in Tables~\ref{t5} and~\ref{t6}.
Although we show the form factors in
Table~\ref{t3n} only at a maximum recoil point $q^2=0$,
we use the form factor at $q^2=M^2_F$ obtained from~\cite{CJBc} for the
corresponding nonleptonic $B_c\to (D_{(s)},\eta_c,B_{(s)})M_F$
decays.

In Table~\ref{t5} we show the nonleptonic decay widths of the
$B_c$ meson for a general value of the Wilson coefficients $a_1$
and $a_2$, whereas in Table~\ref{t6} we give the corresponding
branching ratios (in $\%$) at the fixed choice of Wilson
coefficients~\cite{Gouz}: $a^c_1\;(a^b_1)=1.20\;(1.14)$ and
$a^c_2\; (a^b_2)=-0.317\; (-0.20)$ relevant for the nonleptonic decays
of the $c\; (\bar{b})$ quark. For the lifetime of the $B_c$,
we take the central value $\tau(B_c)=0.46$ ps (i.e. $\Gamma_{\rm
tot}=1.43\times 10^{-12}$ GeV) presented in
PDG~\cite{Data08}. Our
branching ratios for both $b$ and $c$ induced decays listed in
Table~\ref{t6} are generally close to the other quark model
results~\cite{IKS06,EFG03D,EFG03E,HNV,AKN} but differ
substantially from the ones obtained by Refs.~\cite{Gouz,SYDM,CC94}.

The relative size of the branching ratios for various decay modes
may be estimated from power counting of the Wilson
coefficients $a_i$ and the CKM factors with respect to the small
parameter of the Cabibbo angle $\lambda=\sin\theta_C$ in the
Wolfenstein parametrization~\cite{Wol}, e.g. the CKM matrix
elements can be expanded in terms of $\lam$ as $V_{ud}\sim 1$,
$V_{us}\sim\lam$, $V_{ub}\sim\lam^3$, $V_{cd}\sim-\lam$,
$V_{cs}\sim 1$, and $V_{cb}\sim\lam^2$. From Tables~\ref{t5}
and~\ref{t6}, we make the following observations:

(1) The class I decay modes determined by the Wilson coefficient
$a_1$ have comparatively large branching ratios. The CKM favored
$c$ decays such as $B^+_c\to B^0_s\;(\pi^+,\rho^+)$ decays with the
CKM factor $V_{cs}V^*_{ud}\sim\lambda^0$ have branching
ratios of the order of $10^{-2}$ (e.g. $2\%\sim 4\%$ in our model
predictions), which are the most promising class I decay modes
shown in Tables~\ref{t5} and~\ref{t6}.
The CKM-suppressed $c$ decays
such as $B^+_c\to B^0_s K^+$ with $V_{cs}V^*_{us}\sim\lambda^1$
and $B^+_c\to B^0 \;(\pi^+,\rho^+)$ with $V_{cd}V^*_{ud}\sim\lambda^1$,
as well as the CKM-suppressed $b$ decays such as
$B^+_c\to \eta_c\;(\pi^+,\rho^+)$ with $V_{cb}V^*_{ud}\sim\lambda^2$, have
 branching ratios of the order of $10^{-3}$ and should still be accessible
 at high luminosity hadron colliders. However, the branching ratios  of the
 $b\to u$ induced decay modes are too small ${\cal O}(10^{-8}-10^{-6})$
 to be measured experimentally.

 (2) The branching ratios for the class II decay modes
determined by $a_2$ are relatively smaller than those for the
class I decay modes. However, the $B^+_c\to
B^+\;(\bar{K}^0,\bar{K}^{*0})$ decays with
$V_{cs}V^*_{ud}\sim\lam^0$ have branching ratios  of the order
of $10^{-3}$ and these modes should be accessible experimentally.
Of interest is the abnormally small branching ratio of $B^+_c\to
B^+\eta'$ compared to that of $B^+_c\to B^+\eta$. As stated
in~\cite{SYDM}, the reason for such a small branching ratio is not only because
the available physical phase space is too small
but also because there are large destructive interferences between $\eta'_q$ and $\eta'_s$ due
to the serious cancellation between the CKM factors
$V_{cd}V^*_{ud}$ and $V_{cs}V^*_{us}$.

(3) The class III decay modes involve the Pauli interference.
Taking into account the negative value of $a_2$ with respect to
$a_1$, one can see that the class III decay modes shown in
Table~\ref{t5} should be suppressed in comparison with the
cases in which the interference is switched off.
In order to test the effects of the
interference, one may put the widths in the form of
$\Gamma=\Gamma_0 + \Delta\Gamma$, where
$\Gamma_0=x_1a^2_1 + x_2a^2_2$
and $\Delta\Gamma=za_1a_2$, and then compute the
$\Delta\Gamma/\Gamma_0$ (in $\%$) as done in~\cite{Gouz}.
Our absolute values of $\Delta\Gamma/\Gamma_0$ obtained
from
the linear (HO) model are 34.0 (34.0)$\%$ for $B_c^+\to D^+D^0$,
37.3 (42.4)$\%$ for $B_c^+\to D^+_sD^0$, 18.4 (15.1)$\%$ for
$B_c^+\to \eta_c D^+$, and 20.2 (18.9)$\%$ for $B_c^+\to \eta_c
D^+_s$, respectively. This indicates that the interference is the
most significantly involved in the $B_c^+\to D^+_sD^0$ decay
compared to others. In particular, the $B^+_s\to D^+_sD^0$ and
$B^+_c\to D^+_s\bar{D}_0$ decay modes have been proposed
in~\cite{FW00,Mas92,IKP,KisJPG} for the
extraction of the CKM angle $\gamma$ through amplitude relations.

\section{Summary}
In this work, we have studied the exclusive nonleptonic $B_c\to
(D_{(s)},\eta_c,B_{(s)})M$ decays where the final state $M$ mesons
are factored out in the QCD factorization approach.
The inputs used to obtain their branching ratios were the weak form
factors for the semileptonic $B_c\to (D_{(s)},\eta_c,B_{(s)})$
decays in the whole kinematical region  and the unmeasured weak
decay constants obtained from our previous LFQM
analysis~\cite{CJBc,CJ1,CJ2,Choi07}.
For the measured values of decay constants, we use the
central values extracted from the experimental
measurements~\cite{Data08,Artuso,Cleo09}.

Our predictions for the branching ratios are summarized in
Tables~\ref{t5} and~\ref{t6} and compared with other
theoretical results. Overall, the class II decay modes have more
discrepancies among the theoretical models than the class I and
III decay modes do.
The upcoming experimental measurements of the
corresponding decay rates can examine
various theoretical approaches.
The most promising measurable decay modes
appear to be the CKM favored $c$ decays such as
$B^+_c\to B^0_s\;(\pi^+,\rho^+)$ decays. It is thus expected that
the dominant contribution to the $B_c$ total rate comes from the
$c$ induced decays. The more $c$ induced 
$B_c\to VP$ and $B_c\to VV$ decay modes
seem to deserve further consideration.

 \acknowledgments The work of H.-M.Choi was supported by the
Korea Research Foundation Grant funded by the Korean
Government(KRF-2008-521-C00077) and that of C.-R.Ji by the U.S.
Department of Energy(No. DE-FG02-96ER40947).

\end{document}